# High Intensity Secondary Beams Driven by Protons

John Galambos (SNS), Mei Bai (BNL), and Sergei Nagaitsev (FNAL)
August 21, 2013

## 1. Introduction

As part of the Intensity Frontier effort within the 2013 Community Summer Study, a workshop on the proton machine capabilities was held (High Intensity Secondary Beams Driven by Proton Beams) April 17-20, 2013 at Brookhaven National Laboratory in Upton, NY. The agenda and presentations are available on the workshop web-page at: http://www.bnl.gov/swif2013/. Attendees are listed in Appendix A. Primary aims of the workshop were to understand: 1) the beam requirements for proposed high intensity proton beam based measurements; 2) the capabilities of existing world-wide high power proton machines; 3) proton facility upgrade plans and proposals for new facilities; 4) and to document the R&D needs for proton accelerators and target systems needed to support proposed intensity frontier measurements.

The workshop was focused around five guiding questions:

1) What are the particle physics anticipated needs and the requirements for secondary beams, i.e. neutrino, kaon, muon, neutron, etc.
    a. Particles' energy
    b. Particles' flux
    c. Temporal and spatial characteristics
    d. Purity / contamination constraints

2) What are the proton beam requirements to meet the above secondary beam requirements? Are there any overlaps?
3) Can existing accelerator facilities accommodate proton beam requirements in Question 1?
4) What new facilities or upgrades to existing facilities are needed to meet requirements in Question 1?
5) What accelerator and target R&D is required for new facilities and for upgrades of existing facilities?

In summary, this document addresses the above five questions in the following manner:
1. The particle physics requirements on secondary beams are presented in Table 1 and Section 2.
2. The proton beam requirements are also presented in Table 1, which clearly shows that there are many overlaps in proton beam requirements. Both proton beam power on target of 1MW as well as the flexibility in proton beam duty factor are requested.
3. Compatibility of existing facilities proton beam capabilities with the proton beam requirements are listed in Table 2 and Section 3. It is apparent from the workshop discussions that the next generation of particle physics experiments will require capabilities beyond the existing ones, as described in Section 3.
4. Proposed upgrades and new capabilities to meet the requirements are listed in Table 2. Among the proposed, are Project X and Daeδalus (both in the US), as well as SPL and CSNS that are abroad.
5. Discussions on the accelerator and target R&D, both short-term (project specific) and long-term, are reported in in Section 4.

## 2. Particle Physics Anticipated Needs and Requirements: Secondary and Primary Beams

The beam parameters needed for a variety of measurements of interest in the intensity frontier were presented including neutrino, kaon, muon, neutron, and proton EDM measurements. The purpose of this working group is not to prioritize experiments, but rather to describe the primary proton beam requirements needed to facilitate these experiments.

The primary outcome from this portion of the workshop is to identify the proton beam requirements for the different proposed intensity frontier measurements. These requirements are described in Table 1 for the different proposed experiments. The listed primary proton beam requirements are generally not based on detailed calculations but are rather general guidelines. Nonetheless this can be used as a rough guide for matching the desired measurements with existing or proposed proton facilities.

## 3. Primary Beam Capabilities: Existing, Upgrades and Planned

The major proton accelerator facilities were represented at the workshop. U.S facilities included FNAL, BNL, LANL, and SNS. International representation included CERN, J-PARC, PSI, ISIS, and TRIUMF. New and proposed proton accelerator facility representation included ESS, CSNS, Daeδalus, and Project X. Additionally, most existing facilities have upgrade plans to varying degrees. The machine capabilities are summarized in Table 2 below, broken into existing capabilities and new/upgrade proposals. The upgrade plans listed in Table 2 are separated into near-term planned upgrades with some active development, and longer-term plans that are being discussed internally at the institution, but will require significant development. These machine capabilities are characterized in the same manner as the proton machine requirements listed in Table 1, to facilitate identifying possible applications to existing machines or upgrades. Generally, high proton beam powers (~ MW) are desired. The beam energy requirement is typically ≥ 1 GeV while the beam time structure requirements vary significantly, depending on the measurements and specifics of the measurement conditions (e.g. underground detector or not).

In addition to the MW facilities designed for intensity driven experiments, there were also two accelerator-based proposals for precision measurements, including a storage-ring-based proton electric dipole moment measurement and spin physics programs that requires polarized proton beam at the FNAL Main Injector.

## Existing Accelerators

Accelerators with current and anticipated future programs of intensity frontier experiments include FNAL, J-PARC and CERN. All these facilities have accelerator upgrade plans for these ongoing and future experiments (see Table 2).

Regarding the possibility of other existing machine applications, the LANL linac is undergoing a refurbishment and in a few years may have an 800-MeV, 800-kW, 100 Hz x 1 ms proton beam available for potential users. The cold neutron applications may be able to use this beam structure, as well as others.

Another possible application of existing accelerators for desired measurements is for lower energy muon and electron neutrino studies. In particular MW proton beams, with energy <~3 GeV with very low duty factor are desired (to minimize background events from cosmic sources). The J-PARC and SNS neutron sources are possible applications (duty factors < $10^{-4}$). Issues include space for a detector near the neutron production target and the level of background neutrons. ISIS has been used for this purpose in the past, and still has a cavern 10 m from their TS-1 target.

The BNL AGS has accelerated protons to 28 GeV with up to $72 \times 10^{12}$ protons per pulse for slow extraction (55 kW), and up to $55 \times 10^{12}$ protons per pulse (85 kW) for fast extraction. This capability is still available. By replacing the Booster with a 1.2 GeV linac, and an increase of the AGS repetition rate from 0.3 to 2.5 Hz an average power of 1 MW would be accessible, however considerable development is required to establish a complete concept.

## Proposed Accelerators

Proposed new accelerators for intensity frontier applications include Project X, Daeδalus, and nuSTORM which are described below. Of the three, Project X has the broadest scope. In addition, several existing accelerators and construction projects have plans for additional intensity frontier activities. The J-PARC facility has plans for both short-term accelerator upgrades and longer-term efforts aimed at μ-e conversion, $g_\mu$-2/$\mu$EDM, slow extraction >> 100 kW, hyper nuclei an exotic hadron factory and a neutrino factory proton driver. Also the Chinese Spallation Neutron Source (CSNS) has upgrade plans for a long baseline neutrino facility based on post acceleration of a portion of the CSNS beam to a 4 MW, 128 GeV proton source. In the longer term a "megawatt ISIS" may be built at RAL, and in principle could be configured to suit intensity frontier activities.

### nuSTORM

The nuSTORM facility is designed to deliver beams of electron and muon neutrinos from the decay of a stored muon beam with a central momentum of 3.8GeV/c and a momentum acceptance of 10% [1]. The facility is unique in that it will:
- Allow searches for sterile neutrinos of exquisite sensitivity to be carried out;
- Serve the future long- and short-baseline neutrino-oscillation programs by providing definitive measurements of neutrino-nucleon scattering cross sections with percent-level precision for the electron and muon neutrinos.
- Constitutes the crucial first step in the development of muon accelerators as a powerful new technique for particle physics.

nuSTORM is the simplest implementation of the neutrino factory concept. It utilizes protons from the Main Injector, has a conventional (NuMI-like) target station, pion collection with a NuMI-style horn, a pion transport line and a 480m circumference muon decay ring. Although it is foreseen that initial running will be at a power level of 100-150 kW, the target station has been designed for 400 kW, which will allow for higher power levels out of the Main Injector in the era of Project X. No new technology is needed for the facility. The magnet technology needed for the pion transport line and muon decay ring is within current commercial manufacturing capability. The initial detector selected for the neutrino oscillation searches is based iron-scintillator and represents manageable extrapolations from existing (MINOS) detectors. It has been shown through detailed simulation that an exposure of $10^{21}$ protons on target can confirm or exclude the LSND/MiniBooNE results at significance of 10 $\sigma$. In addition, an extensive program of neutrino interaction physics can be done at the near detector hall. The $\nu_e$ interaction physics is a unique capability of the nuSTORM facility and can provide the first measurements of $\nu_e$ cross sections in the range 500 MeV < $E_\nu$ < 4 GeV. The nuSTORM facility can also provide an intense ($10^{10}$/pulse) beam of low-energy muons appropriate for future 6D muon ionization cooling experiments. This beam is available during operations for the neutrino physics program and operates in a completely parasitic mode. Finally the muon decay ring provides the opportunity for accelerator R&D on beam instrumentation and potentially on muon acceleration technology.

## Daeδalus

DAEdALUS is a neutrino research program [2,3,4] based on "decay-at-rest" sources. Pions are produced by interaction of 800 MeV protons on a suitable target (probably composite of graphite and copper), this energy being sufficiently above threshold for good pion yield, and low enough that pions will stop in the target before decaying.

The DAEdALUS configuration consists of three sources of neutrinos as identical as possible, located: Station 1 is 1 to 2 km and 1 MW, for flux normalization; Station 2 is approximately 8 km and ~ 2 MW; and Station 3 - at approximately 20 km and 5 MW. To obtain suitable statistics in a time period commensurate with data-taking of the long-baseline component (5 to 10 years), the proton sources at 800 MeV must be in the megawatt class, delivering currents of 10 milli-amperes on the identical neutrino-generating targets. The cost-performance optimization results in three basically identical cyclotron systems injecting and accelerating molecular hydrogen ($H_2+$) ions. A current of 10 mA is approximately a factor of 5 over the best achieved current at PSI, the world's leading high-power cyclotron today. Using $H_2+$ ions has the potential for alleviating space-charge problems encountered at injection, and extraction of the beam at 800 MeV with a stripper foil minimizes the necessity for clean turn separation at the outer radii, only requiring an extraction channel (for the resulting protons) with sufficiently large momentum acceptance to allow for ions stripped from several overlapping turns.

A cascade of two cyclotrons is planned; an Injector cyclotron with $H_2+$ ions presented along the cyclotron axis and inflected into the central plane. Beam is accelerated to 60 MeV/amu, and is septum-extracted to a separated-sector superconducting ring cyclotron (SRC) to reach the 800 MeV/amu design goal. The 800 MeV protons emerging from the stripper foil spiral inwards, and with suitable placement of the foil can exit the cyclotron approximately 180° away.

Engineering concepts for the DAEdALUS machines build on existing experience. The Injector cyclotron design is a straightforward compact, normal-conducting machine, with an extraction radius of 2 meters. The SRC is very close in size and field to the existing RIKEN SRC cyclotron; outside diameter about 15 meters, and maximum field in each of the 6 sectors is less than 6Tesla.

## Project X

Project X is a high intensity proton facility that will support a world-leading Intensity Frontier research program over the next several decades at Fermilab. When compared to other facilities in the planning stages elsewhere in the world Project X is completely unique in its ability to deliver, simultaneously, up to 6 MW of site-wide beam power to multiple experiments, at energies ranging from 233 MeV to

120 GeV, and with flexible beam formats. Project X will support a wide range of experiments based on neutrinos, muons, kaons, nucleons, and nuclei. A complete concept for Project X has been developed and is documented in the Project X Reference Design Report [5]. A staging strategy has been developed which is comprised of 3 stages with compelling physics capabilities at each stage. The 2013 HEPAP Facilities Subpanel has assessed the science capabilities of Project X as "absolutely central" and the state of development as "ready for construction".

In addition, Project X will lay the foundation for the long-term development of a Neutrino Factory and/or Muon Collider. The Muon Accelerator Program (MAP) has developed a staging plan[6] for these facilities that builds on Stage II of Project X and a detector located at the Sanford Underground Research Facility (SURF). The stages of the long baseline neutrino factory include:
- NuMAX (Neutrinos from Muon Accelerators at Project X): With 1 MW of protons on target and no muon cooling, this facility would store $2 \times 10^{20}$ muons of each species per year and send $8 \times 10^{19}$ neutrinos of each species (electron and muon neutrinos along with their anti-neutrinos) towards the far detector.
- NuMAX+: With ≥3 MW of protons on target and 4D muon ionization cooling, this facility would store $1.2 \times 10^{21}$ muons of each species per year and send $5 \times 10^{20}$ neutrinos of each species towards the far detector.

## 4. Accelerator R&D Needs and Plans

### Project / facility specific R&D Plans

### Project X Development Program [7]

Project X capitalizes on the very rapid development of superconducting RF technologies over the last 20 years, and their highly successful application to high power H⁻ acceleration at the Spallation Neutron Source at Oak Ridge National Laboratory. As a result of these developments excellent simulation and modeling tools exist for designing the Project X facility with high confidence that performance goals can be achieved, and the primary supporting technologies required to construct Project X exist today.

The Project X Collaboration is engaged in a comprehensive development program aimed at mitigating technical and cost risks associated with construction and operations. The Reference Design provides the context for the R&D program; the primary elements of this program are:
- Accelerator Configuration and Performance Projections
- Front End (0-25 MeV) – Integrated Systems Test (PXIE – Project X Injector Experiment)

- H⁻ Injection
- High Intensity Recycler/Main Injector Operations
- High Power Targets
- Superconducting RF

The overall scope and goals of the Project X development program are based on being prepared for a 2017 construction start. Essentially all elements listed above are required for Project X Stage 1 implementation.

### Compact Cyclotron Designs

The DAEdALUS proposed $H_2^+$ acceleration represents a novel approach to providing a potential means of reduced loss charge exchange extraction from a second stage cyclotron acceleration, and a possible lower cost approach to multi- MW high power applications. A systematic R&D program is being planned to address the many challenges of the required performance. These include:
- ion-source developments to achieve high-brightness cold (devoid of loosely-bound vibrational states) $H_2^+$ beams of at least 50 mA CW.
- injection into test cyclotrons to explore bunching efficiency, space requirements and space-charge dynamics; end-to-end simulations to evaluate beam stability and uncontrolled loss.
- atomic physics experiments of stripping and vacuum cross sections and possibly techniques for Lorentz dissociation of vibrational states in high-field (<25T) magnets in transport line between Injector and SRC.
-  details of the stripping extraction process; engineering prototypes of critical components.

Target systems are also challenging and will require substantial R&D as well. Technical reviews including the world's cyclotron experts are planned at critical phases of the development process.

### Chinese HEP Plans

R&D for a medium baseline super-beam based on an ADS-type linac is being pursued by IHEP and IMP jointly, supported by the China-ADS project. The proton driver for the neutrino super-beam facility will be a new machine, but its design and construction will strongly depend on the technical development from these R&D studies and the 250 MeV ADS experimental facility. The 15-MW target and pion/muon collection are big challenges, and R&D studies are needed and are being planned. An experimental area using a 4 kW CSNS beam has also been planned.

For a long baseline super-beam based on CSNS post acceleration, only fast-ramping superconducting magnets are identified to require R&D for this post-acceleration, perhaps in collaboration with FAIR. The target and neutrino horns need R&D, but at present there is no planning for them.

### BNL R&D Plans

Regarding the BNL 200 MeV Linac, upgrade issues being investigated primarily include instrumentation: loss monitors for low energies, BPMs before the target with low energies (beam partially debunched), laser and other profile monitors.

The high intensity R&D items for the other machines (Booster and AGS) are limited, since there are currently no firm plans to go beyond what has been demonstrated in the past.

### ISIS R&D plans

ISIS has several R&D projects being carried out at present: development of higher current H$^-$ ion sources, construction of a 3 MeV H$^-$ front end test stand incorporating a high-performance beam chopper, a comprehensive program of experimental and theoretical beam dynamics studies on the high-space-charge ISIS ring, a neutronics and engineering upgrade to the existing TS-1 target station to deliver a factor ~2 increase in useful neutrons per incident proton, design and integration of a new higher energy linac for injection into the existing ISIS synchrotron[8], and assessment of a 3 GeV synchrotron delivering several megawatts of beam power for neutron spallation (and possibly neutrino factory) purposes[9].

In addition, a program of neutron instrument construction is currently under way — adding four neutron instruments to the ISIS Second Target Station, a target station that is optimized for cold neutrons [10].

### LANL R&D Plans

The most compelling R&D for LANL is to significantly increase the multi-cell superconducting elliptical cavity gradients as compared with the SNS present performance. A factor of approximately two increase in gradient is required for the proposed 3-GeV (or beyond) beam in the present tunnel. Although no specific plans or program funding are in place to reach this goal at present, exploration is ongoing on the gains possible through optimized cavity shapes, improved cavity fabrication and engineering techniques such as hydro-forming, and alternative materials and/or coatings such as MgB2.

A funded, multi-year upgrade program to modernize and maintain high reliability and availability of the Los Alamos Neutron Science (LANSCE) accelerator complex is underway. Primary upgrades include replacement of the 201.25-MHz and 805-MHz RF systems and upgraded controls. In addition to these upgrades, LANSCE Risk Mitigation also includes development of: 1) improved beam position and phase monitors needed to turn on and tune up the beams and 2) near-real-time multi-particle simulations linked to the machine control system (EPICS) and potentially to optimization algorithms designed to more quickly turn on, tune up, and minimize losses in high-intensity machines.

**PSI R&D Plans**

The focus on accelerator development is increased availability since the typical 90% can be improved (the PSI light source has 98% availability). This involves work on electrostatic elements and auxiliary systems that typically cause outages; on the longer term development of new amplifiers/resonators for the injector cyclotron and new high power collimators behind the meson production target; these measures will allow higher powers than 1.4 MW, possibly up to 1.8 MW

Regarding the neutron performance, the target cannot be improved much more; an upgrade program for a better moderator is underway, better signal to noise performance at the experiments (shielding) and maybe better neutron guides; new instruments are proposed, maybe an additional instrument building on the north side.

Ultra-cold neutrons: the PSI experiment should show a very high ultra cold neutron (UCN) density according to simulations, but in practice a factor 30
less is observed(!); a goal is to understand this discrepancy and to improve the present performance significantly (possibly related to frozen deuterium moderator).

A new project (HIMB) is for muons out of the spallation target; close to the target high rates of surface muons are predicted. However, capture of these muons and transport are unclear and will be studied in the next years.

**General Long-term R&D Plans**

**Superconducting Radio Frequency Acceleration**

Most modern high power proton facilities, including those in operation, construction, and planning, rely on linacs utilizing superconducting radio frequency (SRF) acceleration. Examples include the SNS, ESS, CSNS, Project X, and SPL. The initial development of SRF technologies emphasized achievable gradient, in order to limit the physical size of the facilities (SNS, CEBAF, ILC). However, many Intensity Frontier applications require the delivery of high duty factor beams; hence, the figure of merit is shifting from high gradient to high $Q_0$. What is required to support future facilities is the development of cavity fabrication and processing techniques that minimize cost, while reliably producing cavities with gradients in the range 15-25 MV/m, at $Q_0$ in excess of $10^{10}$.

### High-quality beam choppers.

Chopping is essential for high-power megawatt-scale H- linacs injecting into rings, and requirements associated with new facilities require an extension of the current state-of-the-art. Although R&D on beam choppers continues, the technology is proving difficult to perfect. The challenge is to develop choppers to provide full beam deflection on a time scale less than the beam micro-bunch spacing, and/or to cleanly transport partially deflected beam. With beam loss requirements typically less than 1 part per million per meter at full energy, this is a non-trivial task. Also high-quality beam choppers are desirable for high-power proton linacs — to enable flexible bunch timing with good extinction, controlled ramp-up, ramp-down, and production of high quality tune-up beams without extraneous partially chopped beam. Project X has recently proposed a broad-band chopper concept in conjunction with a CW linac beam in order to supply particle physics experiments with variable bunch patterns.

### Laser-assisted stripping injection

The standard technique for producing high intensity proton beams in rings is use of charge exchange injection of H- using stripper foils. At higher powers this becomes problematic for foil survivability (heating) and for beam loss due to the inevitable scattering processes during the initial H- passage through the foil and subsequent foil passages as the beam circulates as a proton. At SNS, the highest activation region is at the ring injection point, due to foil-induced losses. Laser assisted stripping has long been proposed as an alternative means of charge exchange injection which avoids the direct interaction of the beam with material, and the associated beam loss. The concept has been proven in principle with a short demonstration experiment (~10 ns) at SNS [[11], [12]]. However, before it can be fully deployed, areas of required development include minimizing the required laser power for the longer time scale applications (including laser light recycling in Fabry-Perot cavities), high stripping efficiencies for beams with realistic transverse and longitudinal phase space distributions, and minimal emittance increase induced by the multiple steps involved in the entire charge exchange process.

### Large Dynamic Range Beam Instrumentation.

Very precise simulations of beam dynamics involving accurate predictions of beam halos are essential for enabling high-power multi-megawatt-scale machines to run without inducing excessive amounts of radioactivity. However, to properly benchmark the simulation codes, reliable measurements of very weak beam halos in the multidimensional beam phase space will be required. Characterizing beam properties in the 6-D emittance phase space, with a large ($10^4$-$10^5$) dynamic range for low energy beams (to simulate the input beam) and higher beam energies (to compare to the simulated output beam) would provide a more solid foundation for low loss designs.

## FFAG Accelerators

Frequency modulated non-isochronous Fixed Field Alternating Gradient (FFAG) applications are proposed as a path for some future higher power proton facility upgrades (e.g. BLAIRR at BNL), with potential cost savings relative to conventional technologies (RCS, super-conducting linac). High power FFAG applications are a significant step beyond present prototype machines. Some intermediate demonstration steps (e.g. 10-100 kW proton machines) are needed to validate this approach.

## Slow Resonant Extraction / CW Operation

Some proposed intensity frontier measurements (e.g. kaon physics) require near CW, $\geq$ 3 GeV proton beams. The primary means of delivering these beams today is slow resonant extraction from rings.  Plans call for delivery of greater than 100 kW (at relatively low beam energies), but the highest slow extraction to date is ~190 kW at Fermilab (400-GeV protons, 3e13 at 0.1 Hz), BNL has delivered ~ 140 kW at the AGS (8e13 protons every ~2.2 sec at ~24 GeV), CERN has delivered ~100 kW and J-PARC has reached 15-20 kW.  The primary considerations on extending this approach are:

- How much beam loss in the extraction process is tolerable?
    - J-PARC operates with ~100 W, and perhaps up to 1kW is possible, but it is difficult to imagine any higher loss.
    - A side question is how much shielding is practical for the specific implementation?
- What extraction efficiency is possible?
    - J-PARC has the highest efficiency of 99.5%. There are R&D plans to increase this efficiency with high beta insertions and low Z electrostatic septum wires to minimize scattering
- What extraction duty factor is possible?
    - J-PARC has reached 0.43

Although slow extraction operation has been ongoing for several decades, operation with over 100 kW appears to be extremely challenging. R&D is ongoing at J-PARC, and these results will be closely followed.

The straightforward path to high power and high duty factor at 3 GeV is via operation with a CW superconducting RF linac such as Project X proposes. This is a direct solution, with a certain likelihood of success. However it does require the expense of a 3 GeV linac.

Isochronous FFAGs (ring cyclotrons) are also good candidates for high power CW. High output energy cyclotrons (3+ GeV) have been designed in the 1980s but never prototyped. Clean extraction requires separated turns and this sets the scale of both the ring size and the energy gain per turn. Scaling from the PSI ring cyclotron, a beam energy of 3.5 GeV can be reached with a ring 20m in diameter and an energy gain of 10 MeV per turn. Thus, to compare with a superconducting linac, the savings due to 300-fold smaller installed RF gradient but same RF power is to be weighed against the cost of 15 or so large superconducting magnet sectors. These sectors would be fairly complicated and would need some development before feasibility and cost can be confirmed.

### Electrostatic Storage Ring

The proposal for proton EDM physics studies calls for use of a storage ring with electrostatic bending and focusing elements (instead of the nominal magnetic dipoles and quadrupoles). The accelerator needs for this application are somewhat unique (low power with very narrow phase-space beam parameters) compared to the other intensity frontier applications discussed at the workshop. The proposed design is conceptual, but there do not appear to be any show-stoppers preventing implementation of electrostatic bending and focusing at the proposed scale (0.7GeV/c for proton, and 3.1GeV/c for muon). The main challenge of this experiment, similar to most high precision EDM experiments, is how to minimize various systematics including long lasting spin coherence to reach a measurement sensitivity of $10^{-29}$e-cm. The present plan is based on existing technology that needs to be shown as viable in an accelerator environment.

## 5. Target R&D Needs and Plans

Target development and implications for high power proton accelerators should not be underestimated. Three of the present high power accelerator facilities (SNS, J-PARC-MLF, and FNAL-MINOS) noted that during recent operational periods, the beam power was limited by target concerns, not the accelerator capability, some for extended periods.

It was noted that the cost of these systems is a non-negligible portion of high power accelerator facilities, and the resources needed to operate them are also significant (remote maintenance, disposal etc.). Plans for new facilities should include the impacts of handling the proposed high powers.

Although high power target facilities have necessarily specific requirements, they present many common challenges [13]. These are:

- Radiation damage
- Thermal shock effects
- High heat-flux cooling
- Radiation protection and shielding
- Radiation accelerated oxidation effects

- Remote handling
- High intensity beam windows
- Magnets or other devices to focus and direct secondary particles
- Precise predictions of the secondary beam flux
- Machine Protection
- Radiation tolerant instrumentation and monitoring

Overcoming these challenges will require a broad-based program of coordinated R&D activities. At the 2012 Proton Accelerators for Science and Innovation workshop [14] target experts from US and UK institutions identified radiation damage as the leading cross-cutting target facility challenge. In addition it was noted that thermal shock was also a leading issue relevant for both pulsed beam facilities and CW facilities that utilize rotating or flowing targets. These areas are being investigated by engineering groups at many accelerator facilities (Fermilab, STFC-RAL, MSU-FRIB, PSI, GSI, CERN, ESS, ORNL-SNS, J-PARC, LANL), and there has been relatively good communication on these issues among the High Power Target community.

Of the above listed challenges, radiation damage is of special note, due to its complexity, difficulty to evaluate, and length of time to investigate. In addition, simulation plays a lead role in overcoming several of the R&D challenges.

### Radiation Damage R&D

As materials are irradiated, their material properties change due to displacements of atoms in the crystal structure. In addition, transmutation of target atoms generates hydrogen and helium gas which can be detrimental to the material structure. The manner in which the damage manifests in the material properties varies depending upon the material, the initial material structure, the type of radiation, the irradiation dose rate and the irradiation environment (especially irradiation temperature). Many common structural materials, such as stainless steel, can withstand 10 DPA (displacements per atom) or more before reaching end of useful life. However other materials, such as graphite, suffer significant damage at doses as low as 0.1-0.2 DPA [15]. Properties affected by radiation damage include tensile properties, ductility, He embrittlement, thermal and electrical properties, creep, oxidation, and dimensional changes (swelling). In addition, many of these effects are annealed above the irradiation temperature. With overlapping parameters and effects, radiation damage is a complex issue that cannot be taken out of context and must be tested at conditions analogous to operating conditions.

Studies have been conducted over the past 60 years to determine irradiated properties and develop radiation damage tolerant materials for use in the nuclear power industry. Unfortunately such data is from lower energy reactor source neutron radiation and not high energy proton radiation. The differences in gas production, dose rate, irradiation temperature and material type are quite significant between the two irradiation environments.

To investigate radiation damage effects in a coordinated way, Fermilab has initiated the RaDIATE collaboration (Radiation Damage In Accelerator Target Environments) to explore radiation damage issues relevant to high power target facilities (http://www-radiate.fnal.gov/). The RaDIATE Collaboration will draw on existing expertise in related fields in fission and fusion research to formulate and implement a research program that will apply the unique combination of facilities and expertise at participating institutions to a broad range of high power accelerator projects of interest to the collaboration. The broad aims are threefold:

- to generate new and useful materials data for the accelerator and fission/fusion communities;
- to recruit and develop new experts who can cross the boundaries between these communities;
- to initiate and coordinate a continuing synergy between these communities.

Initial participating institutions include Fermilab, PNNL, the Materials for Fusion and Fission Power group at University of Oxford, STFC-RAL, and BNL (with significant interest from MSU-FRIB, ESS, CERN, and LANL). The research and development program currently consists of a research program centered at Oxford on radiation damage effects in beryllium (motivated by the use in high power beam windows), a research activity centered at BNL on radiation damage effects in graphite (motivated by the use as neutrino and ion beam targets), and a study on radiation damage effects in tungsten (motivated by the use in spallation sources). The work at BNL on graphite has been ongoing for the past several years with some interesting results [16]. It is expected that the research program incorporate both bulk sample irradiations for traditional tensile testing as well as lower energy ion shallow irradiations to take advantage of recent developments in micro-mechanics testing.

### Simulation and Failure Limits R&D

Target facility design and operation is heavily dependent upon simulation tools. These tools, while robust for most standard applications, can be pushed beyond their limits when applied to high power targets. Energy deposition, radiation damage (DPA), and gas production results from Monte-Carlo simulations such as MARS, must be validated and benchmarked as input to the thermo-mechanical simulations. Thermo-mechanical simulations can be highly non-linear involving elastic-plastic stress waves, strain-rate dependent effects, and phase change. In addition, incorporating inhomogeneous, time varying, radiation damage effects is quite difficult. So the thermo-mechanical simulations must be also tested and validated. Finally, classically defined failure limits for structural materials (elastic yield limit) may be too conservative for target applications. Failure modes in the unique high power target environment must be explored and characterized to fully utilize and develop novel target materials.

To achieve the above, controlled and instrumented in-beam tests must be performed. CERN has recently begun operation of a facility for such tests, called HiRadMat [17]. Thus far, several materials for the LHC collimator program have been tested along with a novel target material (tungsten powder). Fermilab is designing a test with beryllium (for low-z neutrino targets and beam windows) to run at HiRadMat when it starts operation again in late 2014.

### Secondary Beam Modeling, Optimization, and Measurement

There remain challenges of being able to precisely predict, optimize, and measure the flux of secondary particles produced from a target.

To optimize the flux there are typically arrays of focusing devices, moderators, baffles, collimators, and ducts to deliver the needed beam to the experimental location. These items need experimental permutation and validation to maximize the useful flux, and to actually build devices that can operate in the intense beam environments envisaged.

Predicting the secondary production from a target and through the beam delivery system is quite limited in terms of precision. Dedicated experiments to measure hadron production from various materials with various energy primary and secondary beams are needed to improve the ability to predict secondary production. Software tools and empirical models or parameterizations need to be developed to leverage particular hadron production measurements into arbitrary potential beam configurations.

Furthermore, beamlines include instrumentation that lies within and near the secondary beam to measure the production of secondaries, verify that it conforms to predictions, and monitor it for changes. This instrumentation must measure and survive very high-density particle fluxes; further innovation will allow accommodation of the disparate goals of higher intensity and greater precision.

### The Project X Integrated Target Station (PX-ITS) Notional Concept

The above listed challenges associated with high power target facilities are all present at varying levels in a spallation source proposed as part of the Project X Experimental Facilities, the PX Integrated Target Station (PX-ITS). Because it is envisioned to be a "day one" facility for the first stage of Project X (1 GeV, 0.9 MW), development of PX-ITS will demand that all the target R&D challenges be addressed in a timely manner, thus making it a good test bed for state of the art high power target technology and methods.

The vision for PX-ITS is to provide a high intensity neutron source to power a program of particle physics and nuclear energy relevant experiments simultaneously. The PX-ITS notional concept is currently under development and described in the Broader Impacts of Project X Spallation Source for Irradiation Testing [18] It draws upon expertise from the high energy physics, nuclear physics, and nuclear energy (fusion and fission) communities and is organized by staff at PNNL.

Requirements for PX-ITS will be driven by the physics that can be carried out using neutrons generated initially by a one-megawatt proton beam.  The current experiments being considered for phase I of Project X with a spallation target system include neutron-antineutron oscillation, an UCN neutron source, general neutrino physics, advanced neutron EDM, nuclear EDM, neutrino DAR physics, and muon spin rotation. Some of these experiments can yield insight into physics not currently reachable at the high-energy end using colliders. In addition, and of equal significance, is the inclusion of irradiation volumes and systems for material irradiations relevant to fission and fusion power.

Multiple areas of R&D will be required to carry out the development of the PX-ITS. These areas include radiation transport calculations, shielding analysis, thermal hydraulics, structural analysis, remote handling, utility flow analysis (water, heavy water, and cryogenics systems), target development, etc.  To maintain the current schedule associated with Project X, this development must start as soon as possible.

The next step is to organize a follow-on workshop to the first PX-ITS workshop (then called Energy Station) to discuss the requirements of the physics to be performed [19].  These requirements will drive the further development of the PX-ITS.

As an example, during the development of the Spallation Neutron Source (SNS), many important lessons were learned which can be applied to the PX-ITS.  The most important of these include the absolute necessity of efficient remote handling and the development of methods whereby the cost can be reduced to an absolute minimum while maintaining approved requirements and schedule.

The SNS was developed for a specific purpose. That purpose was to generate room temperature and cryogenic neutrons that would satisfy up to 24 neutron scattering instruments while keeping its ability to be maintained and serviced.  In contrast, PX-ITS should be developed so that it is flexible and reconfigurable to accommodate and optimize yield for multiple experiments and irradiation environments. Test areas and targets/moderator configurations are envisioned to be modularized to allow for extensive re-configuration. Even the shield pile itself could be reconfigurable to accommodate future expansion or modification.

**Target R&D Recommendations**

The following is recommended:

- Continue to support and develop a broad-based program of high power target R&D activities utilizing the capabilities and expertise of the global high power target and nuclear materials communities, including:
    - The RaDIATE R&D program
    - Development of high energy proton irradiation facilities (such as BLAIRR at BNL)
    - High intensity beam on material testing (such as at HiRadMat)
    - Efforts to validate/benchmark simulation methods
    - Development of high intensity beam windows
    - High heat-flux cooling studies

- Develop high power target capability as a core competency at accelerator facilities which have megawatt class target facilities on their 10 year horizon
- Implement a program of hadron production measurements that will address the entire parameter space of particle production relevant for precise calculation of secondary beam fluxes.
- Begin conceptual design activities of the Project X Integrated Target Station (PX-ITS) to spur on high power target development for next generation target facilities, starting with initiation of a series of PX-ITS workshops (including experts from the high energy physics, nuclear energy/materials, nuclear physics and high power target communities) to:
    - Define stage 1 physics/irradiation requirements
    - Iterate on conceptual design layouts
    - Assess regulatory and environmental impacts
    - Explore site options
    - Perform scoping calculations for target systems
    - Determine any additional critical R&D needs

It was noted that the cost of these systems is a non-negligible portion of high power accelerator facilities, and the resources needed to operate them are also significant (remote maintenance, disposal etc.). Plans for new facilities should include the impacts of handling the proposed high powers.

Several spallation neutron source facilities have R&D programs on targets in place (e.g. at ISIS on very compact target configurations based on tantalum-coated tungsten [20]). While these programs may not be aimed primarily at targets for particle physics experiments, nevertheless there is considerable overlap between the physics and engineering for these two purposes that should be taken advantage of.

Table 1. **Next Generation Requirements for the Frontier Capabilities**

| Particle beam | Secondary / tertiary beam requirements | | | | | Proton Beam Requirements * | | | Comments |
|---|---|---|---|---|---|---|---|---|---|
| | Integrated Flux | Purity | Energy | Spatial characteristics | Timing characteristics | Energy | Power | Timing Characteristics | |
| **Photons** | | | | | | | | | |
| **Conventional Super beams:** | | | | | | | | | |
| **Muon neutrinos Kate** | | < 1% electron neutrinos | 0.5-5.0 GeV | Pulsed-horn forward narrow beams | Duty factor < 10$^{-3}$ | >~ 9 GeV | >1MW | Duty factor < 10$^{-3}$ | Classical super-beam technique for both short and long baselines. |
| **Anti-Muon neutrinos Kate** | | < 5% electron neutrinos | 0.5-5.0 GeV | Pulsed-horn forward narrow beams | Duty factor < 10$^{-3}$ | | >1MW | Duty factor < 10$^{-3}$ | Classical super-beam technique for both short and long baselines. |
| **Muon neutrinos** | | Stored muon beam | 3.8 GeV | Forward narrow beams | | 60-120 GeV | 100-400kw | Duty factor < 10$^{-6}$ | Short baseline neutrino factory beams. |
| **Electron neutrinos** | | Stored muon beam | 3.8 GeV | Forward narrow beams | | 60-120 GeV | 100-400kw | Duty factor < 10$^{-6}$ | Short baseline neutrino factory beams. |
| **Muon neutrinos** | | Stored muon beam | 4-6 GeV | Forward narrow beams | | 3-15 GeV | >1MW | Duty factor < 10$^{-3}$ | Long baseline neutrino factory beams. |
| **Electron neutrinos** | | Stored muon beam | 4-6 GeV | Forward narrow beams | | 3-15 GeV | >1MW | Duty factor < 10$^{-3}$ | Long baseline neutrino factory beams. |
| **Pion DAR neutrinos:** | | | | | | | | | |
| **Muon neutrinos** | | | 0-53 MeV: antinumus, | Isotropic, point-source (10 | Duty factor < 10$^{-3}$ | < 3 GeV | >1MW | Duty factor | Proton duty factor applicable for near surface |

|  | Secondary / tertiary beam requirements | | | | | Proton Beam Requirements * | | | |
| --- | --- | --- | --- | --- | --- | --- | --- | --- | --- |
|  |  |  | 30 MeV: numus | m –~ 10 km) |  |  |  | < $10^{-3}$ | detectors, not needed for deep detector ; < 1% nuebar contamination |
| **Electron neutrinos** |  |  | 0-53 MeV | Isotropic, Point-source | Duty factor < $10^{-3}$ | < 3 GeV | >1MW | Duty factor < $10^{-3}$ |  |
| **Kaon DAR neutrinos:** |  |  |  |  |  |  |  |  |  |
| **Muon neutrinos** |  | < 1% electron neutrinos | 235 MeV | Isotropic, point source | Duty factor < $10^{-3}$ | > 3 GeV | >1MW | Duty factor < $10^{-3}$ |  |
|  |  |  |  |  |  |  |  |  |  |
| **Positive Muons for CLFV (B. Bernstein)** | $10^{18}$-$10^{19}$ | >95% | 29.8 MeV surface beam, could go lower | Polyethylene or similar stopping target | CW, microstructure > 500 MHz. | ~1 GeV | 1MW | As CW as possible | Scaling to x10 statistics over MEG at 1 GeV and 3e7 second live. Two experiments, so this much for each |
| **Positive muons for EDMs (Yannis S.)** | $5 \times 10^{16}$ |  | 500 MeV |  | 300 ns |  |  | High rep rate as possible, up to 50 kHz |  |
| **Positive muons for muonium osc. (B. Bernstein)** | $1 \times 10^{15}$ |  | 29.8 MeV surface beam, could go lower | Polyethylene or similar stopping target |  | 1 GeV | 1 kW | 100 kHz | Scaling existing experiment for x100 in statistics and 5 muon lifetimes = 14,841; # of protons in old expt based on hep-ex 9807011 and private data from MEG, large uncertainty |
| **Negative** | $10^{22}$-$10^{23}$ | >95% | As low as | 1-10cm | 0.5-10 MHz of | 1 GeV | 1 MW |  | 6 yr run based on Mu2e |

|  | Secondary / tertiary beam requirements | | | | Proton Beam Requirements * | | | |
|---|---|---|---|---|---|---|---|---|
| **muons for CLFV (B. Bernstein)** | | | possible - 5-70 MeV/c | transverse stopping on thin foils. | <10-100 nsec pulses. | | | | method; FFAG-based calculation immature |
| | | | | | | | | | |
| **Positive Kaons** S. Kettel | $10^{16}$ | >70% at stopping target | 400-600 MeV/c | 1-10cm transverse incident on stopping target. | CW, microstructure > 50 MHz. | 3 GeV | 100kW -1 MW | | |
| **Low energy Neutral Kaons** L. Littenberg | $10^{16}$ | $<10^{-4}$ neutrons per beam kaon outside of beam solid angle | 300-1000 MeV/c | 20-100 μSR neutral beam | 20-50 MHz of <50 psec pulses | > 3 GeV | | | 1.e-7 hadron halo requirement |
| **High energy neutral Kaons** | $10^{16}$ | $<10^{-4}$ neutrons per beam kaon outside of beam solid angle | 1.5-15 GeV | 5 μSR neutral beam | CW | | | | |
| **Cold Neutrons (Y. Kamyshkov)** | $2.47 \times 10^{23}$ neutrons to the exp. Acceptance (2-3 year operation) | Minimal content of gammas, fast neutrons > 1 MeV, and charge particles in the beam | < 1 eV from initial moderator  < 100 neV for UCN | Horizontal beam perpendicular to the proton beam; luminous source area dia 30 cm or area 710 cm$^2$; Focusing equipment access to the cold source starts from the | cold neutron flux: typical ~ 8.75 x $10^{12}$ n/cm$^2$/s/sterad/ MW  CW cold beam, with usage duty factor 80-90% due to proton-beam-pulse-time suppression; | 0.8-2.5 GeV kinetic (1.3 GeV is optimum) | > 0.8 MW | pulsed beam with duty factor 10-20%, e. g. for pulse width 1 ms with repetition 100- | CW operation possible, but less preferred |

| | | | | distance 1 m from the luminous source surface and has opening area 1 x 1 m² (~ 0.66 sterad) | measurement time 2x10⁷ s/yr for 3 yrs. | | | 200 Hz | |
|---|---|---|---|---|---|---|---|---|---|
| **Neutrons EDM (A. Young)** | (1-5)x10²⁰ n/cm²/SR | γ, fast neutron, and charged particles < signal neutron | < 25 meV | Source < 30 cm (2–10)x10¹² n/cm²/s/SR over large solid angle. | Pulsed preferred, CW possible. (2–10)x10¹² n/cm²/s/SR | 0.8 – 2.5 GeV | 200-500 kW | Quasi CW | Cryogenics will be critical |
| | | | | | | | | | |
| **Inclusive charm** | | | | | | | | | |
| **Inclusive Bottom** | | | | | | | | | |
| **Isotopes** | | | | | | | | | |
| **Polarized Protons (Y. Semertzidis)** | 10¹⁵ integrated | > 99% > 80% polarization | 4x10¹⁰ p/20 minutes | | 2 Pulses /20 minutes | 233 MeV, 0.7 GeV/c | 1 mW | < 250 ns base length | $(dp/p)_{rms}$=2.e-4, $\varepsilon_{norm-95\%}$: hor=2 mm mrad vert= 6 mm mrad |
| **High Energy polarized protons (A. Krisch)** | | | | | 1 pulse /1 min. pulse length 5 sec with 10e13 protons | 60-120GeV | | 5 sec per minute | Polarization > 60% |

Table 2. Proton accelerator capabilities: existing, planned upgrades and new facilities.

|  | Energy | Power | Timing Characteristics | Accelerator Type | Comments | Intensity Frontier Application |
|---|---|---|---|---|---|---|
| **Present Capabilities** | | | | | | |
| LANSCE area A | 800 MeV | 80-120kW | 120 Hz | linac | | |
| LANSCE isotope production line | 100 MeV | 250 kW | 40 Hz | linac | Isotope production | |
| ISIS Present | 800 MeV | 200 kW | 0.5 µs x 40 Hz to TS-1, 0.5 µs x 10 Hz to TS-2 | 70 MeV H⁻ linac + RCS | 160 kW to TS-1, 40 kW to TS-2, 30 MeV/c muons from 1-cm-thick graphite target in proton beam line to TS-1 | ≤240 MeV/c muons from pion decay in flight for MICE from Ti target in synchrotron. Cavern within 10 m of TS-1 target — neutrinos? |
| SNS Existing | 935 MeV | 1 MW | 60 Hz x 700 ns | 1 GeV linac + accumulator Ring | parasitic use only - beam on neutron target | Muon and electron DAR neutrinos |
| SNS Existing | 935 MeV | 50-100 kW | 60 Hz x 1 ms | | waste beam from Ring Injection | |
| FNAL Linac present | 400 MeV | ≤12 kW | 15 Hz x 50 µs | H-, NC linac | | Injection into Booster and beam to MTA |
| FNAL Booster present | 8 GeV | 40 kW | 7 Hz x 1.6 us | RCS | | Injection into Main Injector + short baseline muon neutrinos and |

|  | Energy | Power | Timing Characteristics | Accelerator Type | Comments | Intensity Frontier Application |
|---|---|---|---|---|---|---|
|  |  |  |  |  |  | antineutrinos |
| FNAL MI present | 120 GeV | 400 kW | 9.4 us every 2.2 s | RCS | used for neutrino applications | Long baseline muon neutrinos and anti-neutrinos |
|  |  |  |  |  |  |  |
| BNL -BLIP - present | 200 MeV | 25-30 kW | 6.67 Hz x 440 us (5 ns u spacing) | linac | Isotope production + Rad damage |  |
| BNL Booster | 2 GeV | 30 kW | 6.67 Hz | RCS |  |  |
| BNL AGS | 28 GeV | 100 kW | 0.333 Hz | RCS | slow extraction - seconds, fast extraction < 1 us |  |
|  |  |  |  |  |  |  |
| PSI | 590 MeV | 1.3 MW | CW, 50MHz | 2 stage cyclotron | 400 kw to muon target + 900 kW to SINQ neutron scattering facility |  |
|  |  |  |  |  |  |  |
| J-PARC linac | 181 MeV | 20 kW | 25 Hz x 0.5 ms |  |  |  |
| J-PARC RCS | 3 GeV | 300 kW | 25 Hz x 1 us (2 bunches) | 181 MeV linac + RCS | For MLF (Muon facility + neutron scattering) |  |
| J-PARC MR (FX) | 30 GeV | 240 kW | 0.4 Hz x 5 us (8 bunches) | 3 GeV into RCS | For T2K experiment |  |
| J-PARC MR (SX) | 30 GeV | 15 kW | 0.17 Hz x 2 s (debunched) | 3 GeV into RCS | For Hadron Experimental Facility |  |
|  |  |  |  |  |  |  |
| TRIUMF | 60-520 MeV | 100 kW | CW, 23 MHz | cyclotron | 4 simultaneous extraction lines |  |
|  |  |  |  |  |  |  |

|  | Energy | Power | Timing Characteristics | Accelerator Type | Comments | Intensity Frontier Application |
|---|---|---|---|---|---|---|
| **CERN PS** | 14 GeV/c | 45 kW | 1.2 s cycle length, 2.3x10$^{13}$ ppp | | | |
| **CERN SPS operation** | 400 | 470 kW | 4.4 s cycle length, 4200 bunches, 1.05x10$^{10}$p/bunch, 5 ns spacing | linac + 2 stage RCS | CERN Neutrinos to Grand Sasso operation | |
| **CERN SPS** | 400 | 565 kW | 4.4 s cycle length, 4200 bunches, 1.3x10$^{10}$p/bunch, 5 ns spacing | linac + 2 stage RCS | CERN Neutrinos to Grand Sasso operation - record | |
| | | | | | | |
| **Planned Upgrades** | | | | | | |
| | | | | | | |
| **FNAL Boost + PIP (3 years)** | 8 GeV | 90 kW | 15 Hz | | used for neutrino applications + muon physics | Muon neutrinos and antineutrinos; negative muons for CLFV and g-2 |
| **FNAL MI + ANU < 1 year** | 120 GeV | 700 kW | 9.4 us every 1.33 s | | used for NOVA + SX at a few kW | Muon neutrinos and antineutrinos |
| **FNAL MI + ANU < 1 year** | 80-120 GeV | 700 kW | 9.4 us every 0.8 - 1.33s | | by 2020 | Muon neutrinos and antineutrinos |
| | | | | | | |
| **BNL BLIP + upgrade (< 5 yrs)** | 200 MeV | 50-60 kW | 6.67 Hz x 880 us (5 ns u spacing) | linac | Doubling of pulse length | |
| **BNL BLIP post upgrade - BLAIRR (5-10 years)** | > 1 GeV | 250-300 kW | 6.67 Hz x 880 us (5 ns u spacing) | linac | existing linac + FFAG, some beam is needed for isotopes, target R&D, .. | |

|  | Energy | Power | Timing Characteristics | Accelerator Type | Comments | Intensity Frontier Application |
|---|---|---|---|---|---|---|
| **SNS in 4 years** | 1 GeV | 1.4 MW | 60 Hz x 700 ns | linac + accumulator | Planned power increase, no major upgrades | Muon and electron DAR neutrinos |
| **SNS + 2nd target [Ref. 21] (10 years)** | 1.3 GeV | 2.5-3 MW | 60 Hz x 700 ns | linac + accumulator | Energy increased 30%, current increased 50% | Muon and electron DAR neutrinos |
| **LANSCE + Refurbishment (2017)** | 800 MeV | 800 kW | 120 Hz x 625 us | linac | 100 Hz is available for other applications | |
| **ISIS Upgrade , 10–20 years** | 800 MeV | 400 kW | 0.5 μs x 45 Hz to TS-1, 0.5 μs x 5 Hz to TS-2 | 180 MeV H⁻ linac + RCS | 360 kW to TS-1, 40 kW to TS-2, 30 MeV/c muons from graphite target in proton beam line to TS-1 | |
| **J-PARC linac** | 400 MeV | 130 kW | 25 Hz x 0.5 ms | | | |
| **J-PARC RCS** | 3 GeV | 1 MW | 25 Hz x 1 us (2 bunches) | 400 MeV linac + RCS | For MLF (Muon facility + neutron scattering) | |
| **J-PARC MR (FX)** | 30 GeV | 750 kW | 0.75 Hz x 5 us (8 bunches) | 3 GeV into RCS | For T2K experiment | |
| **J-PARC MR (SX)** | 30 GeV | 100 kW | 0.17 Hz x 2 s (debunched) | 3 GeV into RCS | For Hadron Experimental Facility | |

| | Energy | Power | Timing Characteristics | Accelerator Type | Comments | Intensity Frontier Application |
|---|---|---|---|---|---|---|
| **CERN SPS after LHC injector upgrade (2020)** | 400 GeV | 747 kW | 4.4 s cycle length, 4200 bunches, $1.7 \times 10^{10}$ p/bunch, 5 ns spacing | linac + 2 stage RCS | CERN neutrino to Grand Sasso operation | |
| | | | | | | |
| **Long Term Upgrade Concepts** | | | | | | |
| **BNL AGS Upgrade (> 5-10 yrs)** | 28 GeV | 1 MW | 2.5 Hz | RCS | Requires SCL linac injection at 1.2 GeV | |
| **SNS 3rd target ( >10 years)** | 1.3 GeV | 0.5-1 MW | 10-20 Hz x 1 ms | linac | ~ 0.5-1 MW may be available to an alternate station in a long-pulse mode (60 Hz x 1 ms), dependent on 2$^{nd}$ target station requirements | Cold neutrons, neutron EDM for possible 3$^{rd}$ target station |
| **ISIS Upgrade, 20 years +** | 3 GeV | 1 – 5 MW | ~2 µs pulse x 50 Hz, each pulse consisting of 5-10 pulses ~0.3 µs apart | ~800 MeV H⁻ linac + ~3 GeV RCS | Split between target stations undecided | Conceivably combined with proton driver for neutrino factory? |
| | | | | | | |
| **New Machines** | | | | | | |
| | | | | | | |
| **CERN with low power SPL + high power PS** | 75 GeV | 2 MW | 1 Hz x 148 bunches, $1.1 \times 10^{12}$ p/bunch, 25 ns spacing | linac + accumulator | 5 GeV input Energy tp PS, CERN Neutrions to Pyhasaimi program | |

|  | Energy | Power | Timing Characteristics | Accelerator Type | Comments | Intensity Frontier Application |
|---|---|---|---|---|---|---|
| **CERN with high power SPL + accumulator ring** | 5 GeV | 4 MW | 50 Hz x 6 bunches, $1.7 \times 10^{13}$ p/bunch, 160 ns spacing | linac + RCS | Superbeam user |  |
|  |  |  |  |  |  |  |
| **CSNS- Phase I, 2020** | 1.6 GeV | 100 kW | 400 ns x 25 Hz | 80 MeV linac + RCS | 4 kW proton beam for muon production (neutrino R&D, uSR) |  |
| **CSNS- Phase 2, 2024** | 1.6 GeV | 500 kW | 400 ns x 25 Hz | 230 MeV linac + RCS | 4 kW proton beam for mupn production (neutrino R&D, uSR) |  |
| **China long term** | 1.5 GeV | 15 MW | CW, 15 mA |  | Medium baseline superbeam, |  |
| **China longer term** | 128 GeV | 4 MW | 31 uA, $1. \times 10^{-6}$ duty factor, 1.25 Hz | 2 stage RCS + accumulator ring | Long baseline superbeam |  |
|  |  |  |  |  |  |  |
| **ESS** | 2.5 GeV | 5 MW | 14 Hz x 2.9 ms | pulsed SCL | Potential compressor ring upgrade, for spallation, could also source a neutrino beam (Uppsala proposal). |  |
|  |  |  |  |  |  |  |
| **nuSTORM** | 60-120 GeV | 100-400 kw | $1 \times 10^{-6}$ | 3.8 GeV storage ring with 480m circumference fed by the Fermilab MI (also CERN option) | Ring and degraded beam is useful for muon accelerator R&D applications | Neutrinos from muon storage ring: neutrino cross-sections and sterile neutrino search. |
|  |  |  |  |  |  |  |

|  | Energy | Power | Timing Characteristics | Accelerator Type | Comments | Intensity Frontier Application |
|---|---|---|---|---|---|---|
| **Daeδalus** | 800 MeV/amu | 3 stations (1 + 2 + 5) MW | 1 ms x 200 Hz (relatively arbitrary, 20% DF, on time >= 1 ms) | Beam: H2+ Injector: 60MeV/amu compact-cyclotron(IsoDAR) Main cyclotron: separated-sector superconducting (6T) ring, stripping foil extracted | 5 emA peak current (10 mA peak protons on target). 3 accelerator stations at 1.5(1MW), 8(2MW) and 20(5MW) km from detector, operating in sequence at about 20% | Neutrinos from Decay-at-Rest (DAR) of p+ |
| **IsoDAR** | 60 MeV/amu | 600 kW | CW | Compact cyclotron, (DAEdALUS injector) septum extracted | Prototype Injector cyclotron for DAEdALUS, mounted underground close to ~1 kton detector | Electron antineutrinos from DAR of 8Li |
| **Project-X** | | | | | | |
| Conventional Superbeams | | | | | | |
| | 60-120 GeV | 2.3 MW | Duty Factor 1E-5 | Synchrotron | Stage-3 | Muon Neutrinos and Anti-Neutrinos |
| Muon Storage Ring Beams (NuMAX) | | | | | | |

|  | Energy | Power | Timing Characteristics | Accelerator Type | Comments | Intensity Frontier Application |
|---|---|---|---|---|---|---|
|  | 3 GeV | 1-3 MW | Duty Factor 1E-4 | SC CW Linac + Accumulator Ring | Stage 2, duty factor requires ancillary accumulator/compressor rings | Muon Neutrinos and Electron Neutrinos for long baseline neutrino experiments |
| DAR Neutrinos |  |  |  |  |  |  |
|  | 1 GeV | <1 MW | Duty Factor 1E-4 | SC CW Linac + Accumulator Ring | Stage 1, duty factor requires ancillary accumulator ring | Muon Neutrinos from pions |
|  | 3 GeV | >1 MW | Duty Factor 1E-4 | SC CW Linac + Accumulator Ring | Stage 2, duty factor requires ancillary accumulator/compressor rings | Muon Neutrinos from kaons |
|  |  |  |  |  |  |  |
| Muons |  |  |  |  |  |  |
|  | 1 GeV | <1 MW | CW | SC CW Linac | Stage 1 | Positive Muons for CLFV |
|  | 1-3 GeV | <1 MW | 1 MHz of 50 nsec pulses |  | Stage 1,2 | Negative Muons for CLFV |
| Kaons |  |  |  |  |  |  |
|  | 3 GeV | Up to 750 kW | CW, 20 MHz |  | Stage 2 | Positive Kaons |
|  | 3 GeV | Up to 750 kW | CW, 20 MHz, <50 psec bunch length |  | Stage 2 | Low Energy Neutral Kaons |
| Neutrons |  |  |  |  |  |  |
|  | 1 GeV | 0.2 MW | pulsed beam, 20% DF |  | Stage 1, for timescales>1 usec, power = 1 MW x DF | Cold Neutrons |
|  | 1 GeV | 0.9 MW | CW, 40 MHz microstructure |  | Stage 1 | Cold Neutrons |

|  | Energy | Power | Timing Characteristics | Accelerator Type | Comments | Intensity Frontier Application |
|---|---|---|---|---|---|---|
|  | 1 GeV | 0.2-0.5 MW | pulsed beam, 20% DF - CW |  | Stage 1 | Neutron EDM |
| Polarized Protons |  |  |  |  |  |  |
|  | 0.233 GeV | 1 mW | $(dp/p)_{rms}$=2.e-4, enorm-95%: hor=1.2 mm mrad, vert= 1.2 mm mrad |  | Stage 1, momentum spread to be confirmed | Polarized proton EDM |

## Appendix A: Agenda

The agenda is available on the workshop web-page at:
http://www.bnl.gov/swif2013/

## Appendix B: Attendees

**Conveners**: J.Galambos (SNS), S. Nagaistsev (FNAL), M. Bai (BNL)

**Intensity Frontier Liaison**: R. Tschirhart (FNAL)

**Kaon physics**: L. Littenberg (BNL), S. Kettell (BNL)

**Muon physics**: B. Morse (BNL), B. Bernstein (FNAL)

**Neutrino Physics**: M. Bishai (BNL), B. Zwaska (FNAL)
G. Karagiorgi (Columbia), K. Scholberg (Duke), J. Sptitz (MIT)

**Neutron Physics**: Y. Kamyshkov (U. Tenn.), A. Young (NCSU)

**Proton EDM**: Y. Semertzidis (BNL)

**Muon Collider**: M. Palmer (FNAL)

**Polarized proton**: A. Krisch (U. Mich., video)

**International proton facilities**: D. Findlay (ISIS), M. Seidel (PSI)
T. Koseki (J-PARC), Y. Papaphilippou (CERN)
S. Peggs (ESS), J. Tang (CSNS, video), R. Baartman (TRIUMF)

**US proton facilities**: R. Garnett (LANL), B. Weng (BNL), T. Roser (BNL)
W. Fischer (BNL), D. Trbojevic (BNL), S. Holmes (FNAL)
V.Lebedev (FNAL), I. Kourbanis (FNAL)
B. Kephart (FNAL), N. Solyak (FNAL), M. Plum (SNS)
J. Alonso (LBNL, video)

**Target Experts**: N. Simos (BNL), T. Gabriel (U.Tenn.), H. Kirk (BNL)
K. McDonald (Princeton)

**Machine Protection Expert**: R. Schmidt (CERN)

**HEP lab. representatives**: D. Li (LBNL), C. Adolphsen (SLAC), J. Preble (J-Lab)

| Conveners/liaison | 4 |
| --- | --- |
| Physics needs | 14 |
| Accelerator facilities, international | 7 |
| Accelerator facilities, US | 12 |
| System Experts | 5 |
| HEP lab representatives | 3 |
| Total | 45 |